# BER Measurements in the Evaluation of Operation Correctness of VSAT Modem Traffic Interfaces

Jan M. Kelner, Bogdan Uljasz, and Leszek Nowosielski
Institute of Telecommunications, Faculty of Electronics
Military University of Technology
Warsaw, Poland
{jan.kelner, bogdan.uljasz, leszek.nowosielski}@wat.edu.pl

*Abstract*—This paper presents using bit error rate (BER) measurements to evaluate operation correctness of traffic (input-output) interfaces in modem of very small aperture terminal (VSAT). Such functional tests are carried out, for example, when purchasing communication equipment for armed forces. Generally, available standards do not describe measurement procedures in this area. In this case, accredited laboratories should develop dedicated assessment methodologies. In this paper, we show the methodology for the VSAT modems, which is based on the BER measurements.

## I. Introduction

Purchase of communication equipment by armed forces from a civilian market often requires special procedures, tests or adaptations of this equipment to specific conditions for military operations. Such equipment is usually made to order or supplied by leading global companies. Nevertheless, national regulations regarding introducing the equipment to the army require, e.g., tests confirming its relevant parameters and functionalities.

The "IEC 60835" series of standards applies in general to communication systems. Whereas, the "IEC 60835-3" subseries, i.e. [1]-[12], provide methodologies for measuring typical parameters of terrestrial satellite terminals, including: parameters and characteristics of antennas [2], low noise [3] and high power amplifiers [5], up- and down-converters [4], etc. One of these standards [11] focuses on a very small aperture terminal (VSAT). However, none of them concern the assessment of terminal functionalities.

In this paper, we present the verification methodology of supported data interfaces by an examined VSAT modem. This is so-called functional test. Its result is "binary", i.e., the modem supports (result equals 1) or does not support (result equals 0) an analyzed type of the input-output interface. The basis of this methodology is bit error rate (BER) measurement carried out in accordance with [11][13].

The remainder of this paper is organized as follows. Section II presents a short description of the VSAT modem as an equipment under test (EUT). A test-bed and the functional test methodology are shown in Sections III and IV, respectively. In Section V, sample results for different interfaces are presented. We summarize the paper in section VI.

## II. Equipment under Test

### A. Very Small Aperture Terminal

VSAT is a two-way satellite ground station with a parabolic (dish) antenna, whose diameter is less than 3.8 m. Most antennas have a size from 0.7 to 1.4 m. The average transmission rate for VSAT link is from 4kbit/s to 16 Mbit/s. VSATs use satellite transponders located on geosynchronous or geostationary orbits. VSATs can operate in a point-to-point, mesh, or star topology [14][15].

VSAT consists of the following elements: an antenna, orthomode transducer (OMT), block up-converter (BUC), low-noise block down-converter (LNB), interfacility link cable (IFL), indoor unit (IDU). IDU is a VSAT modem, also called a baseband data processing subsystem with a modulator and demodulator or simply – a baseband subsystem.

In the carried-out tests, we use the satellite terminal shown in Fig. 1.

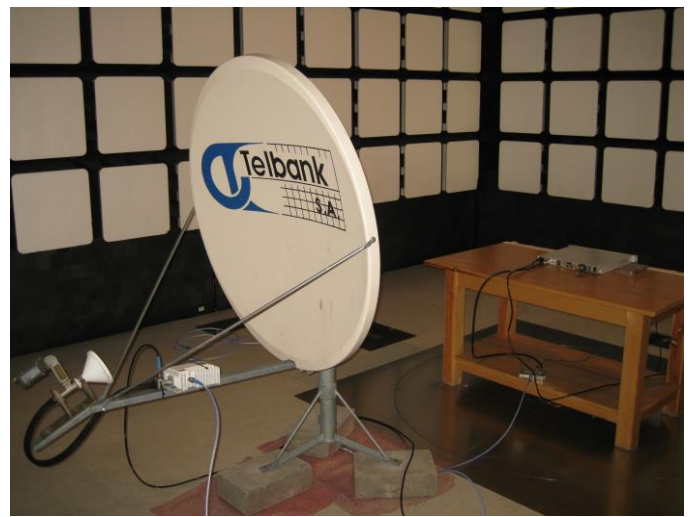

Fig. 1. VSAT used in tests.

*B. VSAT modem*

In the analyzed case, the VSAT modem is EUT. To perform the functional tests in the field of the supported interface verification, the use of other VSAT components is not needed. We carried out tests for the Teledyne Paradise Datacom PD10L modem. The basic parameters of this modem are shown in [16]. Figure 2 presents a front panel of the modem.

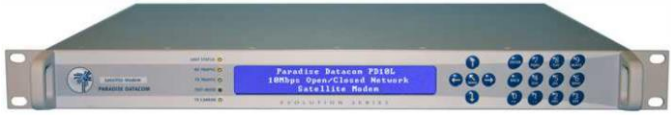

Fig. 2. Teledyne Paradise Datacom PD10L modem [16].

*C. Input-Output Interfaces*

The VSAT modems can be equipped with various traffic interfaces. Currently, Ethernet, i.e., 10/100 Base-T, is most commonly used. Due to the high popularity, most modems are also equipped with G.703 interface.

In [16], the following traffic interfaces are listed:

- standard option: Ethernet (IP traffic on RJ45),
- additional options: G.703 (balanced on EIA530 120Ω or unbalanced on BNC 75Ω female), RS422, X.21, V.35, and RS232 (on EIA530 connector, 25-pin D-type female),
- other optional interfaces: Serial LVDS (25-pin D-type female), Quad E1 G.703 (balanced on RJ45), HSSI (50-pin HD SCSI-2 connector), Eurocom (D/1,D,C,G).

The developed test-bed and measurement procedure provides the opportunity to test several interfaces that are commonly used in communication systems of the Polish Army. They are:

- G.703 [17],
- G.704 [18],
- V.35 [19],
- STANAG 4210 [20],
- 10/100BASE-T,
- 10BASE-FL,
- 100BASE-FX,
- 100BASE-SX.

10BASE-FL and 100BASE-FX/SX, fiber-optic interfaces, 10BASE-T, wired interface, were introduced by the IEEE 802.3-1985 standard [21], while 100BASE-T, wired interface, by the IEEE 802.3-1995 standard [22].

The requirements of traffic interfaces for modems of satellite terminals used in NATO armies are defined in [23].

Prepared test-bed and measurement methodology concern the traffic interfaces listed above.

*D. Interface Converters and Digital Interface Analyzer*

Most commercial VSAT modems do not provide some interfaces that are typical for other communication systems used in the Polish Army, e.g., older standards for optical interfaces, STANAG 4210, or V.35. In this situation, producer or supplier of EUT should equip it with an additional interface converter that will provide its service.

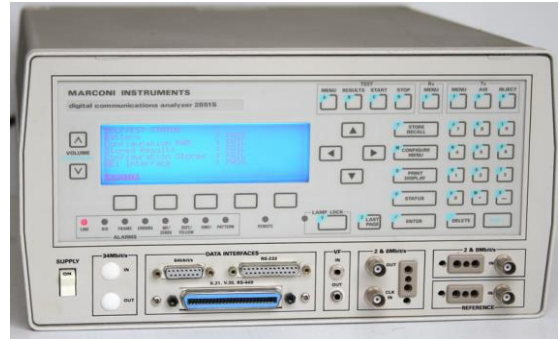

Fig. 3. Marconi 2851S [24].

In test-bed, we use the Marconi 2851S as a digital interface analyzer to measure BER. This device, shown in Fig. 3, is equipped with G.703, G.704, and V.35 (only 512kb/s). Therefore, to provide measurement possibilities for the analyzed types of the traffic interfaces, the test-bed is equipped with appropriate converters:

- Tahoe 284: G.703/G.704 to 10/100BASE-T converter [25],
- Tahoe 235: G.703 to V.35 (to 2Mb/s) converter [26],
- Elektronik Art APP EC100: 10BASE-FL to 10/100BASE-T converter [27],
- Elektronik Art APP EC101: 100BASE-FX/SX to 10/100BASE-T converter [28],
- ZP TEL-KA EUROCOM B/e1: G.703 to STANAG 4210 converter [29].

III. TEST-BED

A scheme of the test-bed is shown in Fig. 4.

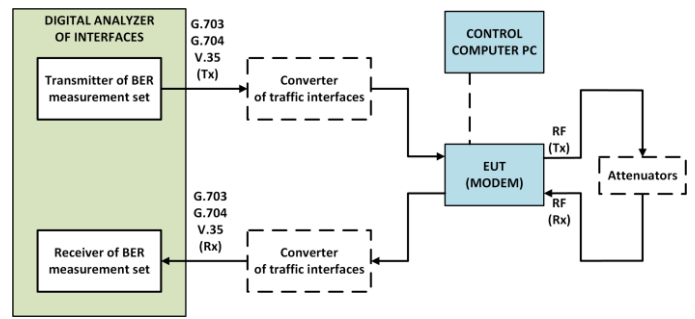

Fig. 4. Scheme of test-bed.

This scheme is developed on the basis of BER measurement methodology presented in [13]. EUT is connected to the interface analyzer, which acts as a BER meter.

This connection may be direct for G.703, G.704, or V.35. Otherwise, the appropriate interface converter should be used. Figure 5 presents the test-bed under studying the G.703 interface.

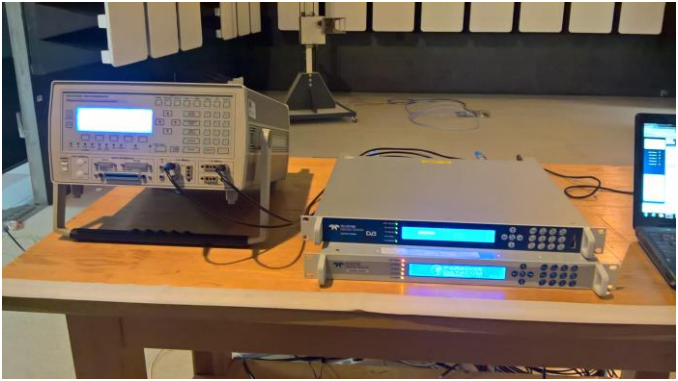

Fig. 5.  Test-bed.

## IV. FUNCTIONAL TEST METHODOLOGY

Procedures for testing the VSATs in a broader scope than described in this paper were developed in the Electromagnetic Compatibility Laboratory [30] of the Military University of Technology [31]. The developed procedures have been accredited by the Military Center for Standardization, Quality and Codification [32]. For quality reasons, all terminal tests except for those requiring communications with space segments, are carried out in an anechoic chamber of the Laboratory. In the general case, the procedure for testing the modem interfaces can be carried out in any room.

This procedure is implemented in the following steps:

a) preparation of the digital analyzer of interfaces - turn on its power supply and leave it in this state for the time necessary to obtain proper stability, and then carry out its self-test;

b) set up the tested satellite terminal modem (EUT) according to its user manual;

c) check if EUT has a proper connector (port) for the analyzed traffic interface type or if EUT has been equipped with an appropriate interface converter; if YES then go to (d) point of the procedure; if NO then the verification of the analyzed interface in the tested modem is negative – in the test report, note that there is no appropriate interface connector and go to (k) point of the procedure;

d) connect EUT to the measuring instrument according to the scheme shown in Fig. 4, selecting the appropriate connector and the type of traffic interface;

e) turn on the EUT power supply and leave it in this state for the time necessary to obtain adequate stability;

f) configure the modem to work with a fixed data transmission speed (bit rate), $B$;

g) tune EUT to the required frequency [33]:

$$f_0 = 0.5\left(f_{max} + f_{min}\right) \quad (1)$$

where $f_{max}$ and $f_{min}$ are maximum and minimum intermediate frequencies of the modem, respectively.

h) set the output (Tx) and input (Rx) parameters for the traffic interface in the digital analyzer of interfaces;

i) measure BER for a selected transmission speed – start the test on the analyzer and then, after the measurement, note the BER result;

j) repeat steps from (f) to (i) for EUT tuning frequencies (in (g) point of the procedure) equal to [33]

$$f_1 = 0.95 f_{max} \quad , \quad f_2 = 1.05 f_{min} \quad (2)$$

k) repeat steps from (c) to (j) for all analyzed types of traffic interfaces.

Analysis of the obtained results of BER measurement consists in checking the condition

$$BER \leq BER_{max} \quad (3)$$

where $BER_{max} = 10^{-5}$ based on [34]. If (3) condition is met then the test is positive, i.e., the modem supports the analyzed interface type.

Based on [11][13], we perform measurements with accuracy $BER_0 = 10^{-8}$. Hence, the time of a single BER measurement is [13]

$$t_0 = 10/(B \cdot BER_0) \quad (4)$$

where $B$ is the selected bit rate (data transmission speed). Table I shows $t_0$ versus $B$.

TABLE I. TIME OF BER MEASUREMENT AS A FUNCTION OF BIT RATE

| Traffic interface | Bit Rate | Time of BER Measurement | |
|---|---|---|---|
| | $B$ (bit/s) | $t_0$ (s) | $t_0$ (h:min:s) |
| V.35 | 64 | 15 625 | 04:20:25 |
| | 128 | 7 813 | 02:10:13 |
| | 192 | 5 208 | 01:26:48 |
| | 256 | 3 906 | 01:05:06 |
| | 320 | 3 125 | 00:52:05 |
| | 384 | 2 604 | 00:43:24 |
| | 448 | 2 232 | 00:37:12 |
| | 512 | 1 953 | 00:32:33 |
| G.703, G.704 (others) | 256 | 3 906 | 01:05:06 |
| | 512 | 1 953 | 00:32:33 |
| | 1 024 | 977 | 00:16:17 |
| | 2 048 | 488 | 00:08:08 |

## V. EXEMPLARY RESULTS

Based on the developed BER measurement methodology, an evaluation of operation correctness of traffic interfaces is carried out for the Teledyne Paradise Datacom PD10L modem shown in Fig. 2. The obtained test results are shown in Table II.

TABLE II. TEST RESULTS FOR TELEDYNE PARADISE DATACOM PD10L

| Traffic interface | BER results | Interface converter is used |
|---|---|---|
| G.703 | $< 10^{-8}$ | NO |
| G.704 | $< 10^{-8}$ | NO |
| V.35 | $< 10^{-8}$ | YES |
| STANAG 4210 | $< 10^{-8}$ | YES |
| 10BASE-T | $< 10^{-8}$ | YES |
| 100BASE-TX | $< 10^{-8}$ | YES |
| 10BASE-FL | $< 10^{-8}$ | YES |
| 100BASE-FX | $< 10^{-8}$ | YES |
| 100BASE-SX | $< 10^{-8}$ | YES |

## VI. CONCLUSIONS

This paper presents the novel evaluation methodology of operation correctness of the VSAT modem traffic interfaces. The basis for the development of the measurement procedure for this functional test are standards of the BER measurements for VSAT. The developed methodology can also be used for other types of satellite modems and communication devices. In addition, the methodology based on the BER measurement can be adapted for other functional tests, such as the evaluation of the bit rate or used coding types.